\documentclass[aps,preprint,showpacs,groupedaddress,floatfix]{revtex4}%
\usepackage{graphicx}
\usepackage{dcolumn}
\usepackage{bm}
\usepackage{amsmath}
\usepackage{amsfonts}
\usepackage{amssymb}%
\newcommand{\be}{\begin{equation}}
\newcommand{\ee}{\end{equation}}
\newcommand{\bea}{\begin{eqnarray}}
\newcommand{\eea}{\end{eqnarray}}

\begin{document}
\title{Covariant theory of particle-vibrational coupling and its effect on the
single-particle spectrum}
\author{E. Litvinova}
\affiliation{Physik-Department der Technischen Universit\"at M\"unchen, D-85748 Garching, Germany}
\affiliation{Institute of Physics and Power Engineering, 249020 Obninsk, Russia}
\author{P. Ring}
\affiliation{Physik-Department der Technischen Universit\"at M\"unchen, D-85748 Garching, Germany}
\date{\today}

\begin{abstract}
The Relativistic Mean Field (RMF) approach describing the motion of
independent particles in effective meson fields is extended by a microscopic
theory of particle vibrational coupling. It leads to an energy dependence of
the relativistic mass operator in the Dyson equation for the single-particle
propagator. This equation is solved in the shell-model of Dirac states. As a
result of the dynamics of particle-vibrational coupling we observe a
noticeable increase of the level density near the Fermi surface. The shifts of
the single-particle levels in the odd nuclei surrounding $^{208}$Pb and the
corresponding distributions of the single-particle strength are discussed and
compared with experimental data.

\end{abstract}

\maketitle

\section{Introduction}

New experimental facilities with radioactive nuclear beams make it possible to
investigate the nuclear chart not only along the narrow line of stable
isotopes but also in areas of large neutron- and proton excess far from the
valley $\beta$-stability. This situation has stimulated enhanced efforts on
the theoretical side to understand the dynamics of the nuclear many-body
problem by microscopic methods. Very light nuclei with $A\leq12$ are studied
by an \textquotedblleft ab initio" approach utilizing bare nucleon-nucleon
interactions of \ two- and three-body character and modern shell-model
calculations based on large scale diagonalization techniques and truncation
schemes show considerable success in predicting properties of somewhat heavier
nuclei. For the large majority of nuclei, however, a quantitative microscopic
description is only possible by density functional theory. These methods are
based on mean-field theory. Although density functional theory can, in
principle, provide an exact description of the many-body dynamics, if \ the
exact density functional is known, in nuclear physics one is far from a
microscopic derivation of this functional. The most successful schemes use a
phenomenological ansatz incorporating \ as many symmetries of the system as
possible and adjust the parameters of these functionals to ground state
properties of characteristic nuclei all over the periodic table. Considerable
progress has been reported recently in constructing such functionals. For a
recent review see \cite{BHR.03}.

Of particular interest are covariant density functionals \cite{Rin.96,VALR.05}
because they are based on Lorentz invariance. This symmetry not only allows to
describe the spin-orbit coupling, which has an essential influence on the
underlying shell structure, in a consistent way, but it also put stringent
considerable restrictions on the number of parameters in the corresponding
functionals without reducing the quality of the agreement with experimental
data. A very successful example is the Relativistic Hartree-Bogoliubov model
\cite{RHB}. It combines a density dependence through a non-linear coupling
between the meson fields \cite{BB.77} with pairing correlations based on an
effective interaction of finite range. A large variety of nuclear phenomena
have been described over the years within this model: the equation of state in
symmetric nuclear matter, ground state properties of finite spherical and
deformed nuclei all over the periodic table \cite{GRT.90} from light nuclei
\cite{LVR.04a} to super-heavy elements \cite{LSRG.96}, from the neutron drip
line, where halo phenomena are observed \cite{MR.96} to the proton drip line
\cite{LVR.04} with nuclei unstable against the emission of protons
\cite{LVR.99}. Recently this model has been also applied very successfully for
the description of excited states, such as rotational bands in normal and
super-deformed nuclei \cite{AKRE.00,ARK.00} and collective vibrations
\cite{MWG.02}. Rotations are treated in the cranking approximation, which
provides a quasi-static description of the nuclear dynamics in a rotating
frame and for the description of vibrations a time-dependent mean field
approximation is used by assuming independent particle motion in
time-dependent average fields \cite{VBR.95}. In the small amplitude limit one
obtains the relativistic Random Phase Approximation (RRPA) \cite{RMG.01}. This
method provides a natural framework to investigate collective and
non-collective excitations of $ph$-character. It is successful in particular
for the understanding of the position of giant resonances and spin- or/and
isospin-excitations as the Gamov Teller Resonance (GTR) or the Isobaric Analog
Resonance (IAR). Recently it has been also used for a theoretical
interpretation of \ low-lying E1-strengths observed in neutron rich isotopes
(pygmy modes) \cite{PRN.03} and for low-lying collective quadrupole
excitations \cite{Ans.05}.

Of course the density functional theory based on the mean field framework
cannot provide an exact treatment of the full nuclear dynamics. It is known to
break down not only in transitional nuclei, where one has to include
correlations going beyond the mean field approximation by treating quantum
fluctuations through a superposition of several mean field solutions, as for
instance in the Generator Coordinate Method (GCM) \cite{RS.80}, but already in
ideal shell model nuclei such as $^{208}$Pb with closed protons and neutron
shells one finds in self-consistent mean field calculations usually a single
particle spectrum with a considerably enhanced Hartree-Fock gap in the
spectrum and a reduced level density at the Fermi surface as compared with the
experiment. It is well known that this fact is connected with the relatively
small effective mass in such models.

Mahaux and collaborators \cite{JLM.76} have shown that the effective mass in
nuclear matter is roughly $m^{\ast}/m$ $\approx0.8$. In finite nuclei it
should be modified by the coupling of the single particle motion to low-lying
collective surface vibrations. This leads, in the vicinity of the Fermi
surface, to an enhancement of $m^{\ast}/m$ $\approx1$. Non-self-consistent
models with the bare mass ( $m^{\ast}/m$ $\approx1$) show indeed a single
particle spectrum with a level density close to the experiment.

Using the quasi-particle concept of Landau theory 
and Green's
function techniques, one can derive a one-body equation for the
single-particle Green's function, which is in principle exact, the Dyson
equation \cite{Mig.67}. It contains a non-local and energy dependent
self-energy, also called mass-operator. The energy independent part of this
operator can be described very well in mean field theory. The most important
origin of an energy dependence is given by the coupling of the single particle
motion to low-lying collective vibrations.

\section{The energy dependent part of the mass operator}

\subsection{One-nucleon motion in the relativistic mean field}

In the relativistic nuclear theory the motion of the nucleons is described by
the Dirac equation
\begin{equation}
\bigl(\gamma^{\mu}P_{\mu}-m^{\ast}\bigr)|\psi\rangle=0,
\end{equation}
where the effective mass is given by
\begin{equation}
m^{\ast}=m+\Sigma_{s} \label{emass}%
\end{equation}
with the scalar part $\Sigma_{s}$ of the mass operator and where the
generalized four-vector momentum operator has the form
\begin{equation}
P_{\mu}=p_{\mu}-\Sigma_{\mu}=\Bigl(i\frac{\partial}{\partial t}-\Sigma
_{0},i\mathbf{\nabla}+\mathbf{\Sigma}\Bigr)
\end{equation}
with the vector part $\Sigma^{\mu}$ of the mass operator
\begin{equation}
\Sigma^{\mu}=(\Sigma^{0},\mathbf{\Sigma}).
\end{equation}
The index '$s$' in the Eq. (\ref{emass}) denotes that the effective mass
is described by the scalar $\sigma$-meson field. In order to characterize ground
state properties the stationary Dirac equation has to be solved:
\begin{equation}
\bigl({\mbox{\boldmath $\alpha$}}(\mathbf{p}-\mathbf{\Sigma})+\beta m^{\ast
}+\Sigma_{0}\bigr)|\psi\rangle=\varepsilon|\psi\rangle. \label{Deq1}%
\end{equation}
In the general case the full mass operator is non-local in the space
coordinates and also in time. This non-locality means that its Fourier
transform has both momentum and energy dependence. Let us assume the
components of mean field to be sums of the stationary local and energy
dependent non-local terms:
\begin{equation}
\Sigma(\mathbf{r},\mathbf{r^{\prime}};\omega)={\tilde{\Sigma}}(\mathbf{r}%
)\delta(\mathbf{r}-\mathbf{r^{\prime}})+\Sigma^{e}(\mathbf{r}%
,\mathbf{r^{\prime}};\omega),
\end{equation}
where all the components of the mass operator are involved:
\begin{eqnarray}
{\Sigma}=({\Sigma}_{s},{\Sigma}^{\mu})\nonumber\\
\tilde{\Sigma}=(\tilde{\Sigma}_{s},\tilde{\Sigma}^{\mu})\nonumber\\
{\Sigma}^e=({\Sigma}_{s}^e,{\Sigma}^{e\mu})
\end{eqnarray}
and index "e" indicates the energy dependence.

The scalar component of the energy-independent mass operator is proportional
to the $\sigma$-meson field:
\begin{equation}
{\tilde{\Sigma}}_{s}(\mathbf{r})=g_{\sigma}\sigma(\mathbf{r}).
\end{equation}
Time-like and space-like components of the local and energy-independent part
of the mass operator ($\tilde{\Sigma}^{\mu}$) are generated by the isoscalar
$\omega$-meson and isovector $\rho$-meson fields $\omega^{\mu},{\vec{\rho}%
}^{\ \mu}$ and Coulomb field $A^{\mu}$:
\begin{equation}
{\tilde{\Sigma}}^{\mu}(\mathbf{r})=g_{\omega}\omega^{\mu}(\mathbf{r})+g_{\rho
}{\vec{\tau}}{\vec{\rho}}^{\ \mu}(\mathbf{r})+e\frac{(1-\tau_{3})}{2}A^{\mu
}(\mathbf{r}),
\end{equation}
where arrows denote isovectors and bold-faced letters indicate vectors in
three-dimensional space. These fields satisfy the inhomogeneous Klein-Gordon
equations:
\begin{align}
(-\Delta+m_{\sigma}^{2})\ \sigma(\mathbf{r})  &  =-g_{\sigma}\rho
_{s}(\mathbf{r})-g_{2}\sigma^{2}(\mathbf{r})-g_{3}\sigma^{3}(\mathbf{r})\\
(-\Delta+m_{\omega}^{2})\ \omega^{\mu}(\mathbf{r})  &  =g_{\omega}j^{\mu
}(\mathbf{r})\\
(-\Delta+m_{\rho}^{2})\ {\vec{\rho}}^{\ \mu}(\mathbf{r})  &  =g_{\rho}{\vec
{j}}^{\ \mu}(\mathbf{r})\\
-\Delta A^{\mu}(\mathbf{r})  &  =ej_{p}^{\mu}(\mathbf{r}),
\end{align}
where the sources are determined by the respective density and current
distributions in a system of A nucleons: the scalar density for $\sigma$-field
\begin{equation}
\rho_{s}(\mathbf{r})=\sum\limits_{i=1}^{A}{\bar{\psi}}_{i}(\mathbf{r})\psi
_{i}(\mathbf{r}), \label{dc1}%
\end{equation}
the baryon current for the $\omega$-field
\begin{equation}
j^{\mu}(\mathbf{r})=\sum\limits_{i=1}^{A}{\bar{\psi}}_{i}(\mathbf{r}%
)\gamma^{\mu}\psi_{i}(\mathbf{r}), \label{dc2}%
\end{equation}
the isovector current for the $\rho$-field
\begin{equation}
{\vec{j}}^{\ \mu}(\mathbf{r})=\sum\limits_{i=1}^{A}{\bar{\psi}}_{i}%
(\mathbf{r})\gamma^{\mu}{\vec{\tau}}\psi_{i}(\mathbf{r}), \label{dc3}%
\end{equation}
and the charge current for the photon-field
\begin{equation}
j_{p}^{\mu}(\mathbf{r})=\sum\limits_{i=1}^{Z}{\bar{\psi}}_{i}(\mathbf{r}%
)\gamma^{\mu}\frac{(1-\tau_{3})}{2}\psi_{i}(\mathbf{r}). \label{dc4}%
\end{equation}
The summation in (\ref{dc1} -- \ref{dc3}) is performed over occupied states in
the Fermi sea, in accordance with \textit{no-sea approximation}, so that the
contribution of the negative-energy states to the densities and currents is neglected.

\subsection{The single-particle Green's function}

In the present work we assume time-reversal symmetry that means the absence of
currents in the nucleus and, thus, vanishing space-like components of
$\ {\Sigma}$. The equation of the one-nucleon motion has the form:
\begin{equation}
\bigl(h^{\mathcal{D}}+\beta\Sigma_{s}^{e}(\varepsilon)+\Sigma_{0}%
^{e}(\varepsilon)\bigr)|\psi\rangle=\varepsilon|\psi\rangle
\end{equation}
or, in the language of Green's functions
\begin{equation}
\bigl(\varepsilon-h^{\mathcal{D}}-\beta\Sigma_{s}^{e}(\varepsilon
)-\Sigma_{0}^{e}(\varepsilon)\bigr)G(\varepsilon)=1, \label{fg}%
\end{equation}
where $h^{\mathcal{D}}$ denotes the Dirac hamiltonian with the
energy-independent mean field:
\begin{equation}
h^{\mathcal{D}}={\mbox{\boldmath $\alpha$}}\mathbf{p}+\beta(m+{\tilde{\Sigma}%
}_{s})+{\tilde{\Sigma}}_{0}.
\end{equation}
We now work in the shell-model Dirac basis $\{|\psi_{k}\rangle\}$ which
diagonalizes the energy-independent part of the Dirac equation:
\begin{equation}
h^{\mathcal{D}}|\psi_{k}\rangle=\varepsilon_{k}|\psi_{k}\rangle.
\end{equation}
In addition we assume in the present work spherical symmetry. In this case the
spinor $|\psi_k\rangle$ is characterized by the set of single-particle quantum numbers
$k = \{(k), m_k \}, (k) = \{t_k, \pi_k, n_k, j_k, l_k\}$ with the radial quantum number 
$n_k$, angular momentum quantum numbers 
$j_{k},m_{k}$, parity $\pi_{k}$ and isospin $t_{k}$:
\begin{equation}
\psi_{k}(\mathbf{r},s,t)=\left(
\begin{array}
[c]{c}%
f_{(k)}(r)\Phi_{l_{k}j_{k}m_{k}}(\vartheta,\varphi,s)\\
ig_{(k)}(r)\Phi_{{\tilde{l}}_{k}j_{k}m_{k}}(\vartheta,\varphi,s)
\end{array}
\right)  \chi_{t_{k}}(t),
\end{equation}
where the orbital angular momenta $l_{k}$ and $\tilde{l}_{k}$ of the large and
small components are determined by the parity of the state $k$:
\begin{equation}
\left\{
\begin{array}
[c]{ccc}%
l_{k}=j_{k}+\frac{1}{2}, & {\tilde{l}}_{k}=j_{k}-\frac{1}{2} &
\mbox{for}\ \ \pi_{k}=(-1)^{j_{k}+\frac{1}{2}}\\
l_{k}=j_{k}-\frac{1}{2}, & {\tilde{l}}_{k}=j_{k}+\frac{1}{2} &
\mbox{for}\ \ \pi_{k}=(-1)^{j_{k}-\frac{1}{2}},
\end{array}
\right.
\end{equation}
$f_{(k)}(r)$ and $g_{(k)}(r)$ are radial wave functions and $\Phi_{ljm}$ is a
two-dimensional spinor:
\begin{equation}
\Phi_{ljm}(\vartheta,\varphi,s)=\sum\limits_{m_{s}m_{l}}({\frac
{{\scriptstyle1}}{{\scriptstyle2}}}m_{s}lm_{l}|jm)Y_{lm_{l}}(\vartheta
,\varphi)\chi_{m_{s}}(s).
\end{equation}
In this basis one can rewrite Eq. (\ref{fg}) as follows:
\begin{equation}
\sum\limits_{l}\bigl\{(\varepsilon-\varepsilon_{k})\delta_{kl}-\Sigma_{kl}%
^{e}(\varepsilon)\bigr\}G_{lk^{\prime}}(\varepsilon)=\delta_{kk^{\prime}},
\label{fg1}%
\end{equation}
where the letter indices $k,k^{\prime},l$ denote full sets of the spherical
quantum numbers mentioned above.

In the next step we represent the exact single-particle Green's function
entering Eq. (\ref{fg}) in the Lehmann expansion. In contrast to the
non-relativistic case, where one has occupied states below the Fermi surface
(hole states $h$) and empty states above the Fermi surface (particle states
$p$) we now have according to the no-sea approximation in addition empty
states with negative energies (antiparticle states $\alpha$). For a detailed
discussion of this point see also Ref. \cite{RMG.01}. Therefore the Lehmann
representation of the Green's function has the form
\begin{equation}
G_{kl}(\varepsilon)=\sum\limits_{h}\frac{\chi_{k}^{h0}\chi_{l}^{h0\ast}%
}{\varepsilon-\varepsilon_{h}-i\eta}+\sum\limits_{p}\frac{\chi_{k}^{0p}%
\chi_{l}^{0p\ast}}{\varepsilon-\varepsilon_{p}+i\eta}+\sum\limits_{\alpha
}\frac{\chi_{k}^{0\alpha}\chi_{l}^{0\alpha\ast}}{\varepsilon-\varepsilon
_{\alpha}+i\eta},
\end{equation}
where $\eta\rightarrow+0$ and the matrix elements are defined as
\begin{equation}
\chi_{k}^{h0}=\langle h|{\hat{\psi}}_{k}|0\rangle,\nonumber
\end{equation}%
\begin{equation}
\chi_{k}^{0p}=\langle0|{\hat{\psi}}_{k}|p\rangle,\nonumber
\end{equation}%
\begin{equation}
\chi_{k}^{0\alpha}=\langle0|{\hat{\psi}}_{k}|\alpha\rangle.
\end{equation}
Here ${\hat{\psi}}_{k}$ is the Dirac field annihilation operator of the state
$k$. $|0\rangle$ denotes the ground state of the subsystem of $N$ particles in
the even-even nucleus in no-sea approximation, i.e. the negative energy states
are essentially empty. The states $|h\rangle$ correspond to the ground state
and to excited states of the subsystem of ($N-1$) particles and $|p\rangle$
are the ground and excited states of the system of ($N+1$) particles,
respectively. Because of the no-sea approximation the negative energy states
$|\alpha\rangle$ are not occupied in $|0\rangle$ and therefore there exist
also states $|\alpha\rangle$ in the ($N+1$) particle system where a level with
negative energy are occupied.

\subsection{The pole structure of the mass operator}

\begin{figure}[ptb]
\begin{center}
\includegraphics*[scale=1.0]{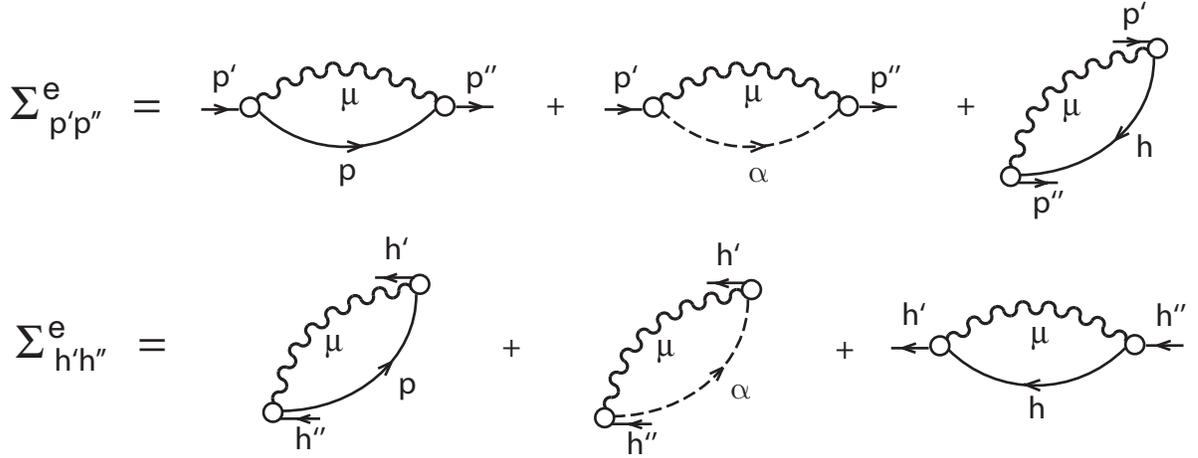}
\end{center}
\caption{The particle $\Sigma^{e}_{p^{\prime}p^{\prime\prime}}$ and the hole
$\Sigma^{e}_{h^{\prime}h^{\prime\prime}}$ components of the relativistic mass
operator in the graphical representation. p, $\alpha$, h are the particle,
antiparticle and hole types of the intermediate states. See text for the
detailed explanation.}%
\label{f0}%
\end{figure}

Let us now define the energy-dependent part of the mass operator (simply
called 'mass operator' in the following). Its matrix elements have the form:
\begin{equation}
\Sigma_{kl}^{e}(\varepsilon)=\int d^{3}rd^{3}{r}^{\prime}~{\psi}_{k}%
^{+}(\mbox{\boldmath $r$})\bigl(\beta\Sigma_{s}^{e}%
({\mbox{\boldmath $r$}},{\mbox{\boldmath $r$}^{\prime}};\varepsilon
)+\Sigma_{0}^{e}({\mbox{\boldmath $r$}},{\mbox{\boldmath $r$}^{\prime}%
};\varepsilon)\bigr)\psi_{l}({\mbox{\boldmath $r$}^{\prime}}). \label{Sigmae}%
\end{equation}
Obviously, on this stage one needs some model assumptions. In the present work
we choose a rather simple particle-phonon coupling model \cite{BM.75} to
describe the energy dependence of $\Sigma^{e}$. Within this model $\Sigma^{e}$
is a convolution of the particle-phonon coupling amplitude $\Gamma$ and the
exact single-particle Green's function \cite{KT.86}:
\begin{equation}
\Sigma_{kl}^{e}(\varepsilon)=\sum\limits_{k^{\prime}l^{\prime}}\int
\limits_{-\infty}^{+\infty}\frac{d\omega}{2\pi i}\Gamma_{kl^{\prime}%
lk^{\prime}}(\omega)G_{k^{\prime}l^{\prime}}(\varepsilon+\omega),
\end{equation}
where the amplitude $\Gamma$ has the following spectral expansion
\begin{equation}
\Gamma_{kl^{\prime}lk^{\prime}}(\omega)=-\sum\limits_{\mu}\Bigl(\frac
{\gamma_{k^{\prime}k}^{\mu\ast}\gamma_{l^{\prime}l}^{\mu}}{\omega-\Omega^{\mu
}+i\eta}-\frac{\gamma_{kk^{\prime}}^{\mu}\gamma_{ll^{\prime}}^{\mu\ast}%
}{\omega+\Omega^{\mu}-i\eta}\Bigr)
\end{equation}
in terms of phonon vertexes $\gamma^{\mu}$ and their frequencies $\Omega^{\mu
}$. They are determined by the following relation:
\begin{equation}
\gamma_{kl}^{\mu}=\sum\limits_{k^{\prime}l^{\prime}}V_{kl^{\prime}lk^{\prime}%
}\delta\rho^{\mu}_{k^{\prime}l^{\prime}}. \label{phonon}%
\end{equation}
$V_{kl^{\prime}lk^{\prime}}$ denotes the relativistic matrix element of the
residual interaction and $\delta\rho$ is the transition density. 
In the present work we use the linearized version of
the model which assumes that $\delta\rho$ is not influenced by the
particle-phonon coupling and can be computed within relativistic RPA. The
linearized version implies also that the energy-dependent part of the mass
operator (\ref{Sigmae}) contains the mean field Green's function ${\tilde{G}%
}(\varepsilon)=(\varepsilon-h^{\mathcal{D}})^{-1}$ instead of the exact Green's
function $G$. So, the equation (\ref{fg1}) becomes linear with respect to $G$.
Since the mean field Green's function is
\begin{equation}
{\tilde{G}}_{kl}(\varepsilon)=\frac{\delta_{kl}}{\varepsilon-\varepsilon
_{k}+i\sigma_{k}\eta},
\end{equation}
where $\sigma_{k}=+1$ if $k$ is an unoccupied state of $p$- or $\alpha$-types
and $\sigma_{k}=-1$ for an occupied $k$ states of $h$-type, the mass operator
$\Sigma^{e}$ takes the form:
\begin{equation}
\Sigma_{kl}^{e}(\varepsilon)= \frac{\delta_{j_{k}j_{l}}\delta_{l_{k}l_{l}}}
{2j_{k}+1}\sum\limits_{\mu,n} \frac{\langle k\parallel\gamma^{\mu(\sigma_{n}%
)}\parallel n\rangle\langle l\parallel\gamma^{\mu(\sigma_{n})}\parallel
n\rangle^{\ast}}{\varepsilon-\varepsilon_{n}-\sigma_{n}(\Omega^{\mu}-i\eta
)}\ . \label{mo}%
\end{equation}
Here we use the notation:
\begin{equation}
\langle k\parallel\gamma^{\mu(\sigma_{n})}\parallel n\rangle=\delta
_{\sigma_{n},1}\langle k\parallel\gamma^{\mu}\parallel n\rangle\ +\ \delta
_{\sigma_{n},-1}\langle n\parallel\gamma^{\mu}\parallel k\rangle^{\ast}.
\end{equation}
Since the indexes $k, l$ and $n$ run through the whole Dirac basis, each state
in (\ref{mo}) can be a particle above the Fermi surface, a hole below the
Fermi surface or a particle in a state with negative energy (antiparticle
state). The graphical representation of the mass operator is given in Fig.
\ref{f0}. We draw the particle and the hole components assuming all the
possible types of intermediate states. Solid line with arrow denotes a
particle (hole) in the Fermi sea, dashed line means a particle in the empty
Dirac sea, weavy line is a phonon propagator, and small circle denotes a
phonon vertex (\ref{phonon}). Time direction is from the left to the right.
\newline One can see from the Eq. (\ref{mo}) that the matrix
$\Sigma_{kl}^{e}$ contains a small number of the off-diagonal elements with
relatively large energy denominators. Additionally, it was shown by explicit
calculations within the non-relativistic approach \cite{RW.73} that it
is justified to use the diagonal approximation:
\begin{equation}
\Sigma_{kl}^{e}(\varepsilon)=\delta_{kl}\Sigma^e_{k}(\varepsilon)
\label{diagonal}%
\end{equation}
with%
\begin{equation}
\Sigma_{k}^{e}(\varepsilon)=\frac{1}{2j_{k}+1}\sum\limits_{\mu,n}%
\frac{|\langle k\parallel\gamma^{\mu(\sigma_n)}\parallel n\rangle|^{2}}{\varepsilon
-\varepsilon_{n}-\sigma_{n}(\Omega^{\mu}-i\eta)}\ . \label{mo1}%
\end{equation}
In analogy with the conventional terminology of non-relativistic approaches,
let us call the intermediate term $n$ 'polarization term' if $\sigma
_{n}=\sigma_{k}$ and 'correlation term' if $\sigma_{n}=-\sigma_{k}$. The
correlation term describes, obviously, the backwards going diagrams in
Feynman's language and corresponds to the ground state correlations caused by
the particle-vibration coupling.

Thus, within the diagonal approximation of the mass operator (\ref{diagonal})
the exact Green's function $G$ is also diagonal in the Dirac basis and the Dyson
equation forms for each $k$ a non-linear eigenvalue equation
\begin{equation}
(\varepsilon-\varepsilon_{k}-\Sigma_{k}^{e}(\varepsilon))G_{k}(\varepsilon)=1.
\label{fg2}%
\end{equation}
The poles of the Green's function $G_{k}(\varepsilon)$ correspond to the zeros
of the function
\begin{equation}
f(\varepsilon)=\varepsilon-\varepsilon_{k}-\Sigma_{k}^{e}(\varepsilon).
\label{zeros}%
\end{equation}
For each quantum number $k$ there exist several solutions $\varepsilon
_{k}^{(\lambda)}$ characterized by the index $\lambda$. Because of the
coupling to the collective vibrations the single particle state $k$ is
fragmented. In the vicinity of the pole $\varepsilon_{k}^{(\lambda)}$ the
Green's function can be represented as follows:
\begin{equation}
G_{k}^{(\lambda)}(\varepsilon)\simeq\frac{S_{k}^{(\lambda)}}{\varepsilon
-\varepsilon_{k}^{(\lambda)}+i\sigma_{k}\eta},
\end{equation}
where the residuum $S_{k}^{(\lambda)}$ has a meaning of the single-particle
(hole) strength of the state $\lambda$ with single-particle quantum numbers
$k$. Differentiation of the equation (\ref{fg2}) with respect to $\varepsilon$
provides the expression for the residua:
\begin{equation}
S_{k}^{(\lambda)}=\Bigl(1-\frac{d\Sigma_{k}^{e}(\varepsilon)}{d\varepsilon
}\mid_{\varepsilon=\varepsilon_{k}^{(\lambda)}}\Bigr)^{-1}. \label{sfac}%
\end{equation}
There are several ways to solve the equation (\ref{fg2}). In the present work
we employ the method which has been used in Ref. \cite{RW.73} to solve the
similar problem in the non-relativistic framework. Since the mass operator of
the form (\ref{mo}) has a simple-pole structure, it is convenient to reduce
the Eq. (\ref{fg2}) to a diagonalization problem of the following matrix:
\begin{equation}
\left(
\begin{array}
[c]{cccc}%
\varepsilon_{k} & \eta_{kn_{1}}^{\mu_{1}} & \eta_{kn_{2}}^{\mu_{1}} & \cdots\\
\eta_{kn_{1}}^{\mu_{1}\ast} & \sigma_{n_{1}}\Omega^{\mu_{1}}+\varepsilon
_{n_{1}} & 0 & 0\\
\eta_{kn_{2}}^{\mu_{1}\ast} & 0 & \sigma_{n_{2}}\Omega^{\mu_{1}}%
+\varepsilon_{n_{2}} & 0\\
\vdots & 0 & 0 & \ddots
\end{array}
\right)  , \label{matrix}%
\end{equation}
where
\begin{equation}
\eta_{kn_{i}}^{\mu}=\frac{\langle k\parallel\gamma^{\mu(\sigma_{n_i})}\parallel n_{i}%
\rangle}{\sqrt{2j_{k}+1}}.
\end{equation}
The eigenvalues of the matrix (\ref{matrix}) are the desired poles
$\varepsilon_{k}^{(\lambda)}$ of the exact Green's function. The structure of
the solution is well known: these eigenvalues lie between the poles of the
mass operator. Eventually, the spectroscopic factors have to be calculated at
the points of these poles according to (\ref{sfac}):%
\begin{equation}
S_{k}^{(\lambda)}=\Bigl(1 + \frac{1}{2j_{k}+1}\sum\limits_{\mu,n}%
\frac{|\langle k\parallel\gamma^{\mu(\sigma_n)}\parallel n\rangle|^{2}}{(\varepsilon
_{k}^{(\lambda)}-\varepsilon_{n}-\sigma_{n}\Omega^{\mu})^{2}%
}\ \Bigr)^{-1}. \label{sfac1}%
\end{equation}

\section{Details of the calculations and discussion}

The matrices (\ref{matrix}) have been diagonalized for the single-particle
states $k_{i}$ of both neutron and proton subsystems belonging to the four
major shells around $N=126$ and $Z=82$. Thus, the eigenvalues with the largest
spectroscopic factors correspond to the single-particle excitations of the
nuclei $^{207}$Pb, $^{209}$Pb, $^{207}$Tl and $^{209}$Bi. In subsection
\ref{ill} we discuss the effect of states with negative energies in the Dirac
sea on the mass operator relying on results obtained within the restricted
particle-phonon space. More realistic results for energies and spectroscopic
factors obtained in an enlarged particle-phonon space are presented and
discussed in subsection \ref{real}. In subsection \ref{other} we compare the
present method with other approaches.

\subsection{Relativistic effects: illustrative calculations}

\label{ill} \ The main interest of the present work is to describe the effects
of complex configurations within the relativistic scheme. Therefore, first we
investigate the contributions of pure relativistic terms to the mass
operator and, hence, their influence on the single-particle spectrum of odd
nuclei. In order to keep the numerical effort within a reasonable limit we
used a restricted particle-phonon space taking into account only the most
collective phonons with spin and parity $J^{\pi}=2^{+},$ $3^{-},$ $4^{+},$
$5^{-},$ $6^{+}$ below the neutron separation energy and a reduced number of
single particle states with positive energy (particles or holes). This enables
one to reduce strongly the number of poles in the mass operator (\ref{mo1}) as
well as the dimension of the matrix (\ref{matrix}). Notice, that since in the
Green's function formalism we stay in the single-particle basis, it is always
possible to vary the number of phonons, and the problem of the completeness of
the phonon basis does not arise at all. In all these calculations we use the
parameter set NL3 \cite{NL3} \ for the Lagrangian.

The numerical results obtained in these investigations are compiled in the
Table \ref{tab1}. For the first shell of neutron levels above ('particle') and
below ('hole') the Fermi level three versions are given: in the version
$ph\alpha$ the index $n$ in Eq. (\ref{mo1}) includes all contributions from
intermediate states above the Fermi level ($p$, with $\sigma_{n}=+1$), below
the Fermi level ($h,$ with $\sigma_{n}=-1$) and in the Dirac sea ($\alpha,$
with $\sigma_{n}=+1$). Version $p\alpha$ (for particles) or $h$ (for holes)
excludes the backward going diagrams (i.e. only states with $\sigma_{n}%
=\sigma_{k}$ are taken into account), and the third version $ph$ does not
contain antiparticle intermediate states in (\ref{mo1}).

\begin{table}[ptb]
\caption{Energies $\varepsilon_{k}^{(d)}$ and spectroscopic factors
$S_{k}^{(d)}$ of the dominant neutron levels in $^{208}$Pb calculated in the
strongly restricted particle-phonon space. $ph\alpha$ denotes full the
calculation, $p\alpha$ ($h$) is the version without backwards going terms, and
$ph$ is the version without contribution of the antiparticle states in the
mass operator (see text for details).}%
\label{tab1}
\begin{center}
\vspace{3mm} \tabcolsep=1.25em \renewcommand{\arraystretch}{1.1}%
\begin{tabular}
[c]{cccccccc}\hline\hline
State $k$ & $\varepsilon_{k}$, MeV & \multicolumn{3}{c}{$\varepsilon_{k}%
^{(d)}$, MeV} & \multicolumn{3}{c}{$S_{k}^{(d)}$}\\
\hline
Particle & \  & $ph\alpha$ & $p\alpha$ & $ph$ & $ph\alpha$ & $p\alpha$ &
$ph$\\
\hline
2g9/2 & -2.50 & -2.85 & -3.14 & -2.88 & 0.89 & 0.92 & 0.89\\
1i11/2 & -2.97 & -2.82 & -3.20 & -2.90 & 0.94 & 0.97 & 0.94\\
1j15/2 & -0.48 & -1.16 & -1.33 & -1.21 & 0.70 & 0.74 & 0.70\\
3d5/2 & -0.63 & -0.96 & -1.05 & -0.98 & 0.93 & 0.94 & 0.93\\
4s1/2 & -0.36 & -0.88 & -0.92 & -0.89 & 0.93 & 0.93 & 0.93\\
2g7/2 & -0.56 & -0.71 & -0.90 & -0.76 & 0.92 & 0.94 & 0.92\\
3d3/2 & -0.02 & -0.35 & -0.42 & -0.37 & 0.93 & 0.93 & 0.93\\
\hline
Hole & \  & $ph\alpha$ & $h$ & $ph$ & $ph\alpha$ & $h$ & $ph$\\
\hline
3p1/2 & -7.66 & -7.67 & -7.40 & -7.70 & 0.96 & 0.98 & 0.96\\
2f5/2 & -9.09 & -8.97 & -8.71 & -9.02 & 0.93 & 0.96 & 0.93\\
3p3/2 & -8.40 & -8.20 & -7.87 & -8.22 & 0.90 & 0.94 & 0.90\\
1i13/2 & -9.59 & -9.30 & -9.07 & -9.36 & 0.90 & 0.92 & 0.89\\
2f7/2 & -11.11 & -10.20 & -9.98 & -10.22 & 0.72 & 0.76 & 0.72\\
(1h9/2)$_{1}$ & -13.38 & -13.32 & -13.23 & -13.34 & 0.52 & 0.47 & 0.53\\
(1h9/2)$_{2}$ & \  & -12.48 & -12.42 & -12.49 & 0.31 & 0.39 &
0.29\\\hline\hline
\end{tabular}
\end{center}
\end{table}In this way, one can see that the effects of ground state
correlations (GSC) caused by the particle-phonon coupling and neglected in the
second version are significant and it is essential to take them into account
in a realistic calculation. On the other hand, the contribution of the
antiparticle subspace to the mass operator is quantitatively not of great
importance. This can be understood by the fact that these configurations
provide large values for the energy denominators in (\ref{mo1}). Thus it is
justified to disregard them in the full calculation. Notice, however, that
version $ph$ does not eliminate the effects of the Dirac sea completely since
the phonon vertices still contain this contribution. As it has been discussed
in Ref. \cite{RMG.01} these terms play an important role in a proper treatment
of relativistic RPA. Otherwise it is not possible to obtain reasonable
properties for the isoscalar modes within RRPA.

\subsection{Realistic calculations in an enlarged space}

\label{real} In this section we neglect the effects of the Dirac sea, i.e. the
intermediate index $n$ in Eq. (\ref{mo1}) runs only over particle states above
the Fermi level and holes below the Fermi level. It has been found in section
\ref{ill} that this is a very reasonable approximation. Since we do not have
to include these contributions, we are now able to enlarge the particle-hole
basis considerably by taking into account particle-hole configurations far
away from the Fermi surface. In this case we increase the collectivity of the
phonons and, consequently, the strength of the particle-vibrational coupling.
The phonon basis was also enriched by including higher-lying modes up to 35
MeV, although these modes are not so important as the low-lying ones.
Vibrations with the quantum numbers of spin and parity $J^{\pi}=2^{+},$
$3^{-},$ $4^{+},$ $5^{-},$ $6^{+}$ were included in the phonon space. One
should keep in mind, however, that in the solution of the RRPA-equations for
the vibrational states besides the usual $ph$-components a large number of
$\alpha h$-components of the Dirac sea was included. Of course, as it is
usually done in the RRPA calculations, both Fermi and Dirac subspaces were
truncated at energies far away from the Fermi surface: in the present work we
fix the limits $\varepsilon_{ph}<100$ MeV and $\varepsilon_{\alpha h}>-1800$
MeV with respect to the positive continuum. The energies and B(EL)$\uparrow$
values of the most collective phonon modes calculated with the parameter set
NL3 are displayed in Table \ref{tab2} together with some experimental data.
\begin{table}[ptb]
\caption{Energies and reduced transition probabilities of the most collective
vibrations in $^{208}$Pb obtained within RRPA and experimental data from
\cite{RS.80}.}%
\label{tab2}
\begin{center}
\vspace{3mm} \tabcolsep=2.50em \renewcommand{\arraystretch}{0.65}%
\begin{tabular}
[c]{ccccc}\hline\hline
& \multicolumn{2}{c}{RRPA \hphantom{abc}} & \multicolumn{2}{c}{Exp.
\hphantom{abc}}\\
J$^{\pi}$ & $\omega$ & B(EL)$\uparrow$ & $\omega$ & B(EL)$\uparrow$\\
& (MeV) & (e$^{2}$fm$^{2L}) $ & (MeV) & (e$^{2}$fm$^{2L}) $\\\hline
&  &  &  & \\
2$^{+}$ & 4.98 & 2.69$\times$10$^{3}$ & 4.07 & 2.97$\times$10$^{3}$\\
\  & 5.84 & 5.82$\times$10$^{2}$ & \  & \ \\
\  & 8.38 & 1.22$\times$10$^{3}$ & \  & \ \\
\  & 12.40 & 4.08$\times$10$^{3}$ & \  & \ \\
\  & 22.96 & 1.08$\times$10$^{3}$ & \  & \ \\
&  &  &  & \\
3$^{-}$ & 2.74 & 7.46$\times$10$^{5}$ & 2.61 & 5.40(30)$\times$10$^{5}$\\
\  & 4.95 & 5.81$\times$10$^{4}$ & \  & \ \\
\  & 7.29 & 5.90$\times$10$^{4}$ & \  & \ \\
\  & 22.27 & 6.12$\times$10$^{4}$ & \  & \ \\
&  &  &  & \\
4$^{+}$ & 4.96 & 1.39$\times$10$^{7}$ & 4.32 & 1.29$\times$10$^{7}$\\
& 6.14 & 5.49$\times$10$^{6}$ & \  & \ \\
& 8.01 & 1.08$\times$10$^{7}$ & \  & \ \\
& 9.10 & 2.67$\times$10$^{6}$ & \  & \ \\
& 11.69 & 3.88$\times$10$^{6}$ & \  & \ \\
& 13.67 & 2.93$\times$10$^{6}$ & \  & \ \\
& 14.26 & 3.45$\times$10$^{6}$ & \  & \ \\
& 18.90 & 2.62$\times$10$^{6}$ & \  & \ \\
&  &  &  & \\
5$^{-}$ & 3.14 & 5.16$\times$10$^{8}$ & 3.19 & 4.62(55)$\times$10$^{8}$\\
& 4.31 & 3.01$\times$10$^{8}$ & 3.71  & 3.30$\times$10$^{8}$ \\
& 5.73 & 1.69$\times$10$^{8}$ & \  & \ \\
& 7.26 & 5.13$\times$10$^{8}$ & \  & \ \\
& 11.14 & 3.39$\times$10$^{8}$ & \  & \ \\
& 15.26 & 1.30$\times$10$^{8}$ & \  & \ \\
& 17.29 & 4.58$\times$10$^{8}$ & \  & \ \\
& 22.87 & 2.32$\times$10$^{8}$ & \  & \ \\
&  &  &  & \\
6$^{+}$ & 4.96 & 4.15$\times$10$^{10}$ & 4.42  & 2.30$\times$10$^{10}$ \\
& 6.19 & 2.09$\times$10$^{10}$ & \  & \ \\
& 6.74 & 1.22$\times$10$^{10}$ & \  & \ \\
& 9.72 & 8.65$\times$10$^{9} $ & \  & \ \\
& 11.88 & 1.10$\times$10$^{10}$ & \  & \ \\
& 27.53 & 1.14$\times$10$^{10}$ & \  & \ \\
& 33.85 & 9.45$\times$10$^{9} $ & \  & \ \\
& 34.85 & 9.27$\times$10$^{9} $ & \  & \ \\\hline\hline
\end{tabular}
\end{center}
\end{table}

As one can see from the Table \ref{tab2}, the characteristics of low-lying
modes obtained in RRPA with the parameter set NL3 are, in general, in
accordance with experimental data. For the lowest $2^{+}, 4^{+}$ vibrations 
the B(EL)$\uparrow$ values are in a good agreement with experimental ones, only the energies are
slighty too high, whereas for the lowest $3^{-}, 5^{-}$ their energies are reproduced rather
well and the B(E3)$\uparrow$ value is to some extent overestimated, and the
first $6^{+}$ state is more collective within RRPA then the observed one.

In the present calculations the phonon space was confined also by a criterion
for the B(EL)$\uparrow$ values: all modes with B(EL)$\uparrow$ values less
than 10\% of the maximal one were neglected for $2^{+},3^{-},4^{+}$, and less
than 20\% \ for higher multipolarities. Nevertheless, the mass operator
(\ref{mo1}) has been calculated in a rather wide particle-phonon space and
therefore the single-particle strength is distributed over many states. The
typical dimension of the matrix (\ref{matrix}) is about two thousand and it
varies depending on the state $k$. As it was mentioned above, contributions of
antiparticle states ($n=\alpha$ in the intermediate sum over $n$ in the mass
operator (\ref{mo1}) ) were excluded because they provide large values of the
energy denominators in (\ref{mo1}). On the other hand, the contributions of
the correlation terms (i. e. the terms with $\sigma_{n}=-\sigma_{k}$ in the
intermediate sum over $n$ in the mass operator (\ref{mo1}) ) have been fully
taken into account since they are found to be quantitatively important and
they compensate to some extent the polarization terms.

The final results of these calculations are compiled in the Table \ref{tab3}.
All the energies are related to the experimental ground states of the
respective odd nuclei surrounding the doubly magic nucleus $^{208}$Pb. The
numbers $\varepsilon_{k}$ denote the RMF single particle energies,
$\varepsilon_{k}^{(d)th}$ are the eigenvalues of the matrix (\ref{matrix})
with the maximal spectroscopic factor, and $\varepsilon_{k}^{(d)exp}$ are the
experimentally observed excitation energies. We display here only the results
for one major shell below and one shell above the Fermi surface because in the
next shells almost all the single-particle levels turn out to be strongly
fragmented due to phonon coupling and it is no longer possible to determine
the dominant levels in these shells, in other words, the concept of Landau
quasi-particles breaks down at energies far away from the Fermi level.

\begin{table}[ptb]
\caption{Energies $\varepsilon_{k}^{(d)}$ and spectroscopic factors
$S_{k}^{(d)}$ of the dominant single-particle levels in odd nuclei of the
$^{208}$Pb surroundings calculated in the wide particle-phonon space. The
experimental data are taken from \cite{RW.73}.}%
\label{tab3}
\begin{center}
\vspace{3mm} \tabcolsep=1.60em \renewcommand{\arraystretch}{0.8}%
\begin{tabular}
[c]{ccccccc}\hline\hline
\  & \  & \multicolumn{3}{c}{Energy, MeV} & \multicolumn{2}{c}{Spectroscopic
factors}\\
Nucleus & State $k$ & $\varepsilon_{k}$ & $\varepsilon_{k}^{(d)th}$ &
$\varepsilon_{k}^{(d)exp}$ & $S_{k}^{(d)th}$ & $S_{k}^{(d)exp}$\\\hline
&  &  &  &  &  & \\
$^{209}$Pb & 2g9/2 & 1.44 & 0.65 & 0.00 & 0.84 & 0.78$\pm$0.1\\
\  & 1i11/2 & 0.97 & 0.66 & 0.78 & 0.88 & 0.96$\pm$0.2\\
\  & 1j15/2 & 3.46 & 2.10 & 1.42 & 0.66 & 0.53$\pm$0.1\\
\  & 3d5/2 & 3.31 & 2.55 & 1.56 & 0.88 & 0.88$\pm$0.1\\
\  & 4s1/2 & 3.58 & 3.02 & 2.03 & 0.92 & 0.88$\pm$0.1\\
\  & 2g7/2 & 3.38 & 2.80 & 2.49 & 0.86 & 0.72$\pm$0.1\\
\  & 3d3/2 & 3.92 & 3.31 & 2.54 & 0.89 & 0.88$\pm$0.1\\
&  &  &  &  &  & \\
$^{209}$Bi & 1h9/2 & -0.79 & -1.24 & 0.00 & 0.88 & 1.17\\
\  & 2f7/2 & 2.37 & 0.93 & 0.89 & 0.77 & 0.78\\
\  & 1i13/2 & 2.78 & 1.31 & 1.60 & 0.61 & 0.56\\
\  & 2f5/2 & 4.36 & 2.73 & 2.81 & 0.60 & 0.88\\
\  & 3p3/2 & 5.64 & 3.64 & 3.11 & 0.56 & 0.67\\
\  & 3p1/2 & 6.39 & 4.89 & 3.62 & 0.37 & 0.49\\
&  &  &  &  &  & \\
$^{207}$Pb & 3p1/2 & 0.29 & 0.31 & 0.00 & 0.90 & 1.07\\
\  & 2f5/2 & 1.72 & 1.39 & 0.57 & 0.86 & 1.13\\
\  & 3p3/2 & 1.03 & 0.89 & 0.89 & 0.86 & 1.00\\
\  & 1i13/2 & 2.22 & 1.73 & 1.63 & 0.81 & 1.04\\
\  & 2f7/2 & 3.74 & 2.34 & 2.34 & 0.64 & 0.88\\
\  & 1h9/2 & 6.01 & 4.59 & 3.41 & 0.36 & 1.10\\
&  &  &  &  &  & \\
$^{207}$Tl & 3s1/2 & 0.13 & 0.40 & 0.00 & 0.84 & 0.95\\
\  & 2d3/2 & 1.23 & 1.32 & 0.35 & 0.85 & 1.15\\
\  & 1h11/2 & 2.19 & 1.91 & 1.34 & 0.78 & 0.89\\
\  & 2d5/2 & 2.86 & 2.04 & 1.67 & 0.68 & 0.62\\
\  & 1g7/2 & 7.02 & 5.73 & 3.47 & 0.22 & 0.40\\\hline\hline
\end{tabular}
\end{center}
\end{table}

The difference between $\varepsilon_{k}$ and $\varepsilon_{k}^{(d)th}$ is the
shift of the single-particle level $k$ caused by the coupling to collective
surface vibrations. Notice, that almost all the levels are moving downwards
providing thus a considerably better agreement with experimental energies then
the pure RMF states. One can see also from this table that the dominant
neutron and proton levels obtained in these calculations have large
spectroscopic factors \ and are therefore rather good single-particle states.
This is in agreement with experiment. Nonetheless the single-particle strength
is distributed over many levels. One can easily understand the origin of these
large spectroscopic factors from the structure of solutions of the
(\ref{fg2}). Since each root lies between the two neighbouring poles of the mass 
operator, only one root can be found in the rather wide window
\begin{equation}
\varepsilon_{h}^{(max)}-\Omega^{\mu(min)}\leq\omega\leq\varepsilon_{p}%
^{(min)}+\Omega^{\mu(min)}, \label{win}%
\end{equation}
i.e -10.11 MeV $\leq\omega\leq$ -1.20 MeV for neutron and -10.75 MeV
$\leq\omega\leq$ -1.06 MeV for proton subsystems, respectively. Therefore, for
the single-particle state $k$ near the Fermi surface other roots turn out to
be far away because the lowest $3_{1}^{-}$ phonon has rather high energy in
magic nuclei, therefore the mixing is weak and the respective spectroscopic
factor (\ref{sfac1}) is close to one. On the contrary, if the state is near
the limits or outside of the window, there are many other roots in the
vicinity, and a strong mixing leads to the appearance of several levels with
comparable strength. It can be easily understood why in open-shell nuclei such
a mixing is much stronger near the Fermi surface: the window (\ref{win}) is
noticeably smaller due to both the smaller gap between the last occupied and
the first unoccupied levels, and much the lower energy of the first 2$^{+}$
phonons (see, for instance, \cite{AK.99}).
\begin{figure}[ptb]
\begin{center}
\includegraphics*[scale=1.0]{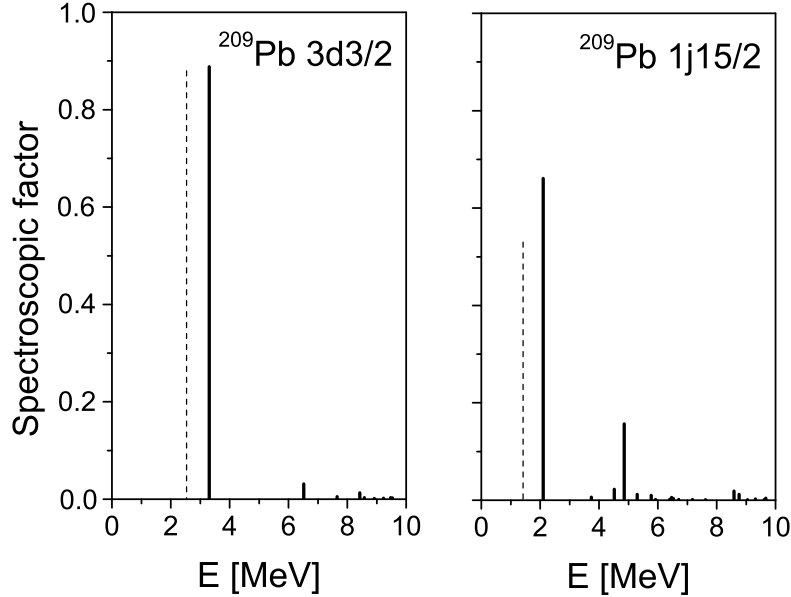}
\end{center}
\caption{Single-particle strength distribution for the $3d_{3/2}$ (left panel)
and $1j_{15/2}$ (right panel) states in $^{209}$Pb obtained in the
calculations (solid lines) and the experimental strengths of the respective
dominant levels (dashed lines).}%
\label{f2}%
\end{figure}\begin{figure}[ptb]
\begin{center}
\includegraphics*[scale=1.0]{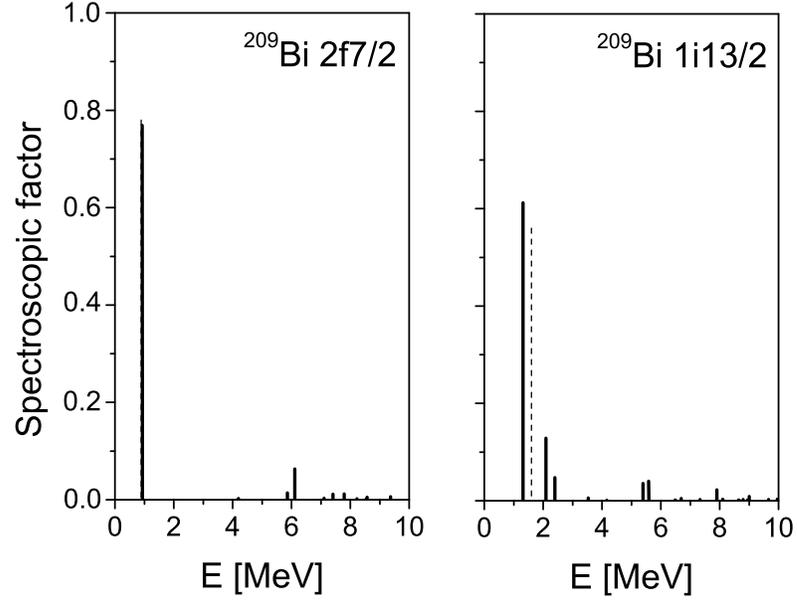}
\end{center}
\caption{The same as in Fig. \ref{f2} but for the $2f_{7/2}$ (left panel) and
$1i_{13/2}$ (right panel) states in $^{209}$Bi.}%
\label{f3}%
\end{figure}\begin{figure}[ptb]
\begin{center}
\includegraphics*[scale=1.0]{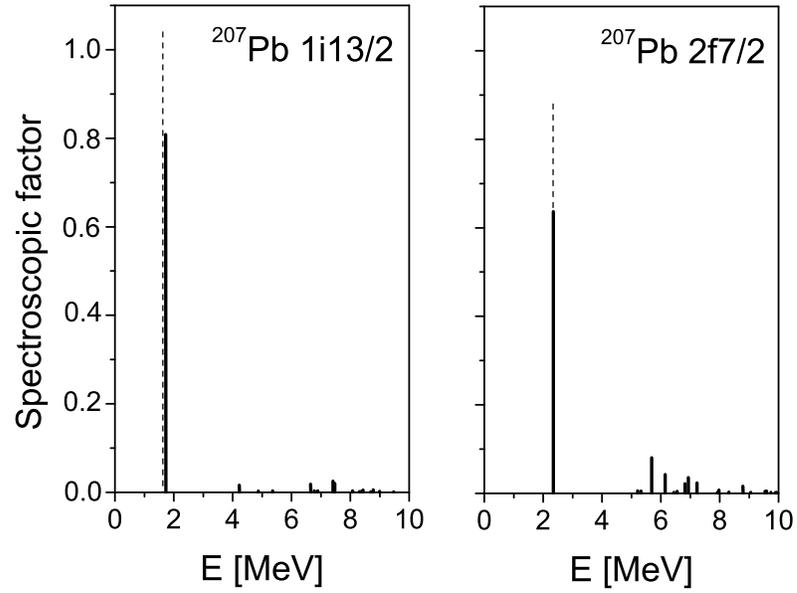}
\end{center}
\caption{The same as in Fig. \ref{f2} but for the $1i_{13/2}$ (left panel) and
$2f_{7/2}$ (right panel) states in $^{207}$Pb.}%
\label{f4}%
\end{figure}\begin{figure}[ptb]
\begin{center}
\includegraphics*[scale=1.0]{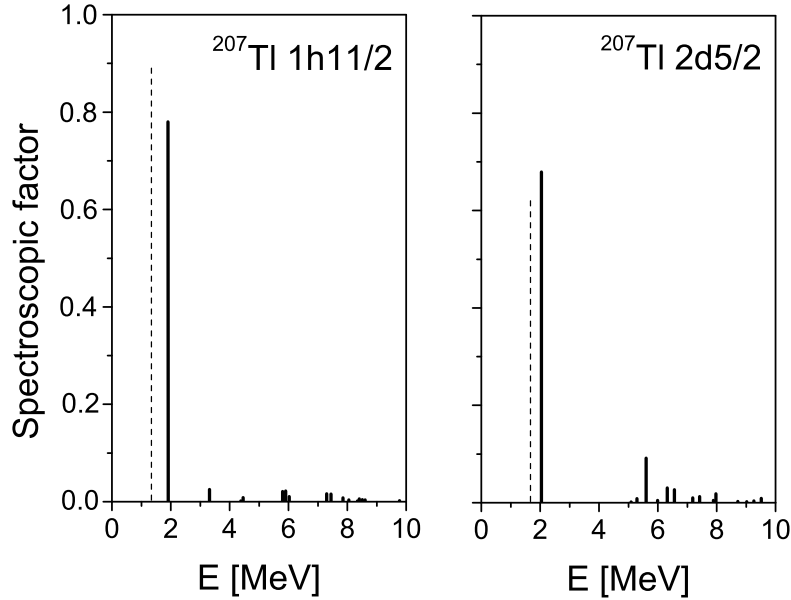}
\end{center}
\caption{The same as in Fig. \ref{f2} but for the $1h_{11/2}$ (left panel) and
$2d_{5/2}$ (right panel) states in $^{207}$Tl.}%
\label{f5}%
\end{figure}\begin{figure}[ptb]
\begin{center}
\includegraphics*[scale=1.5]{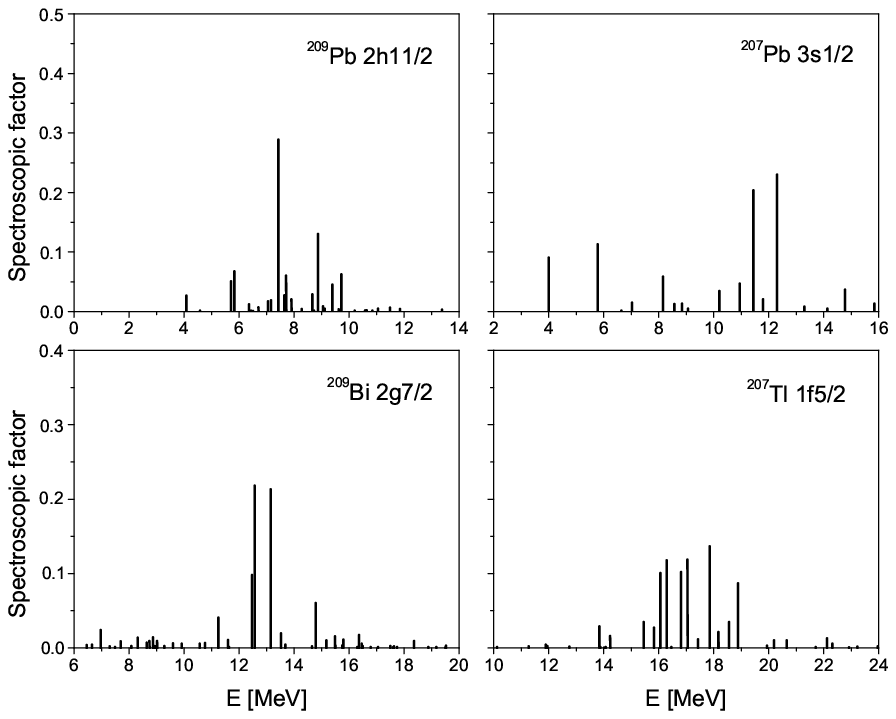}
\end{center}
\caption{The typical strongly fragmented states far from the Fermi surface in
the odd mass nuclei surrounding $^{208}$Pb calculated within RMF with
allowance for the particle-vibration coupling.}%
\label{f6}%
\end{figure}

Nevertheless, some dominant levels near the Fermi surface have noticeably
reduced strength because of the particle-phonon coupling. This is also
confirmed by experiment. One can see such a situation, for instance, for the
$1j_{15/2}$ and the $2f_{7/2}$ neutron states in $^{209}$Pb and $^{207}$Pb,
respectively. Only for the state $1h_{9/2}$ in the $^{207}$Pb we do not find
good agreement: the spectroscopic factor is less then a half of the
experimental value. One also should keep in mind that the experimental
spectroscopic factors depend considerably on the parameters used in the DWBA
analysis. The proton states are found to be somewhat more fragmented then the
neutron states while the dominant levels for the protons above the Fermi
surface are more strongly shifted relatively to the RMF values then in the
those of the neutrons.

The distributions of the single-particle strength for selected levels in
$^{209}$Pb, $^{209}$Bi, $^{207}$Pb, and $^{207}$Tl are represented in the
Figs. \ref{f2}, \ref{f3}, \ref{f4}, \ref{f5}, respectively.

To illustrate the results of these calculations we chose one state of
pronounced single-particle nature (left panels) and one noticeably fragmented
state (right panels) for each nucleus, both from the first major shell
above and below the Fermi surface. As in Table \ref{tab3} all the energies are
related to the ground state energy of the corresponding odd nucleus. As it was
already mentioned, in the present calculations the single-particle strength is
distributed over about two thousand states but most of them are vanishingly
small, so only the states with the strength exceeding 10$^{-3}$ are drown. The
experimental strength of the dominant levels are shown with dashed lines. Some
examples of the strongly fragmented states from the second major shells
above and below the Fermi surface are shown in the Fig. \ref{f6}.

\begin{figure}[ptb]
\begin{center}
\includegraphics*[scale=1.2]{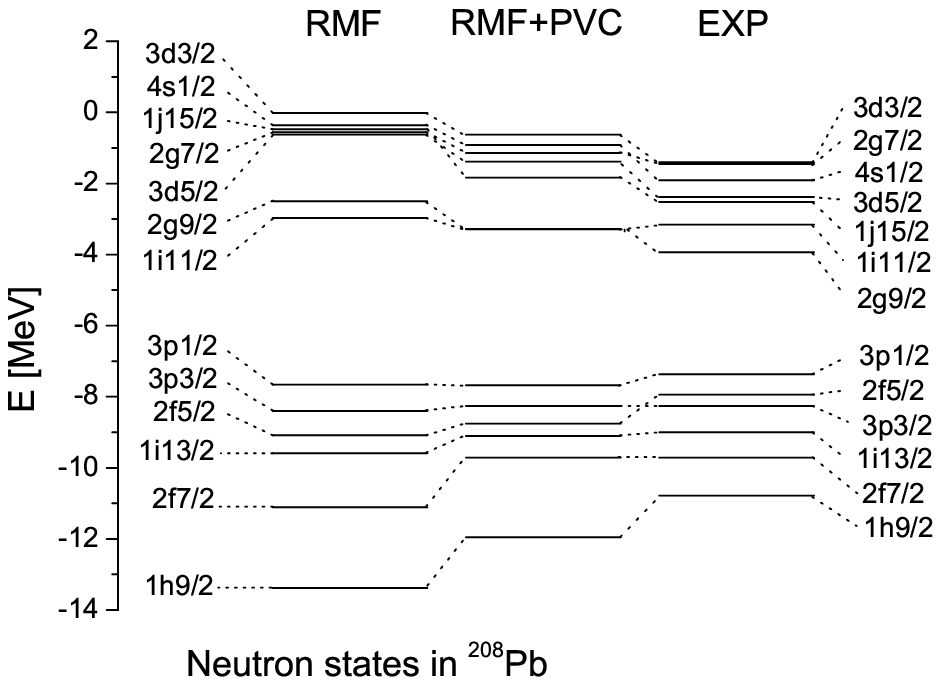}
\end{center}
\caption{Neutron single-particle states in Pb$^{208}$: the pure RMF spectrum
(left column), the levels computed within RMF with allowance for the
particle-vibration coupling (center) and the experimental spectrum (right).}%
\label{f7}%
\end{figure}

\begin{figure}[ptb]
\begin{center}
\includegraphics*[scale=1.2]{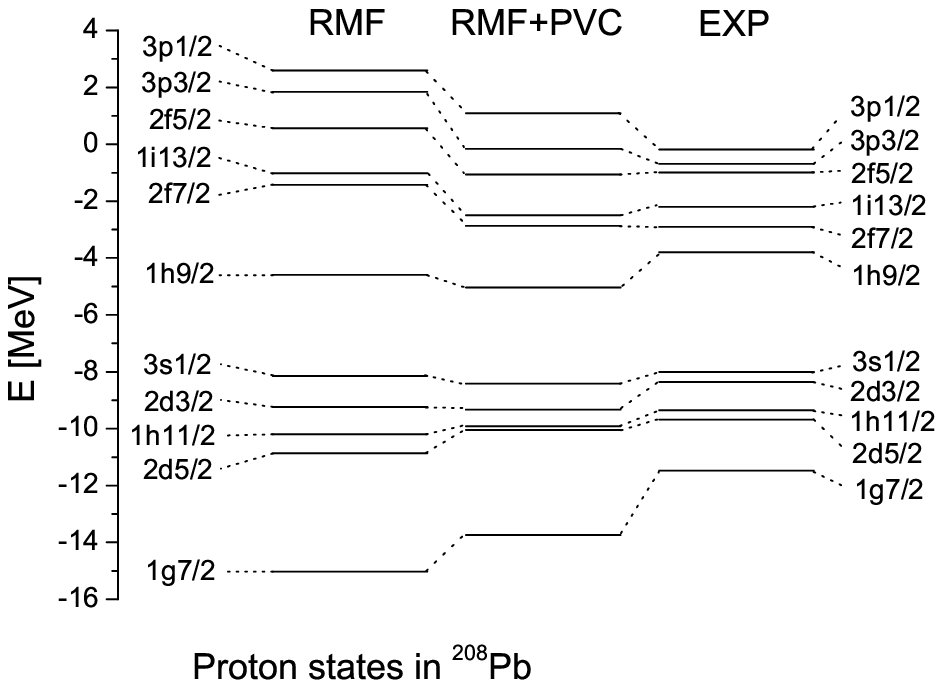}
\end{center}
\caption{The same as in Fig. \ref{f7}, but for proton single-particle states.}%
\label{f8}%
\end{figure}

To illustrate the shifts in the level schemes we show in Figs. \ref{f7} and
\ref{f8} the single-particle spectra for neutrons and protons. The spectra
calculated with the energy-dependent correction (RMF+PVC) demonstrate a
pronounced increase of the level density around the Fermi surface of $^{208}%
$Pb both for neutron and proton subsystems comparatively the pure RMF spectra.
In some cases it turned out to be possible to invert the order of
levels and reproduce the observed sequence as one can see for the$\ 1j_{15/2}$
and the $3d_{5/2}$ neutron states. Another and more important example is the
inversion of the $2g_{9/2}$ and $1i_{11/2}$ neutron states (in the Fig. \ref{f7}
they look coincided) which enables one to reproduce the spin of the $^{209}$Pb ground state.

In order to quantify these results we calculated the average distance between
two levels in the spectrum shown in Figs. \ref{f7} and \ref{f8} (1h9/2 and 1g7/2
states with small spectroscopic factors were excluded from the estimation of the 
neutron and proton spectrum, respectively).
We obtain for the neutrons 1.0 (RMF), \ 0.83 (RMF+PCV) and 0.76 (EXP)  in units of MeV.
This corresponds to a level density of 1.0 (RMF), 1.20 (RMF+PCV) and 1.31
(EXP) in units of MeV$^{-1}$. The level density in the neighborhood of the
Fermi surface is therefore in RMF-calculations by a factor 0.76 smaller than
the experimental value. Taking into account particle-vibrational coupling we
find only a reduction of \ 0.92. \ Assuming an effective mass close to 1 for
the experiment, and taking into account that the level density at the Fermi
surface is proportional to $m^{\ast}/m$, this correspond to an effective mass
$m^{\ast}/m\approx0.76$ for the RMF and $m^{\ast}/m\approx0.92$ for the
RMF+PCV calculations. \ For the protons the situation is similar. From Fig.
\ref{f8} we obtain for the average level distance 1.50 (RMF), \ 1.24 (RMF+PCV)
and 1.06 (EXP) in units of MeV, i.e. the level density is 0.67 (RMF), 0.81
(RMF+PCV) and 0.94 (EXP) in units of MeV$^{-1}$. This corresponds to an
effective mass $m^{\ast}/m\approx0.71$ for the RMF and $m^{\ast}/m\approx0.85$
for the RMF+PCV calculations. We observe that the values for the effective
masses of the protons are slightly smaller than those of the neutrons.

Jaminon and Mahaux \cite{JM.89,JM.90} have discussed in great detail the
concept of the effective mass in the case of RMF theory. On one side one has
the well known Dirac mass%
\begin{equation}
m_{D}=m+\tilde{\Sigma}_{s}(\mathbf{r),}%
\end{equation}
which is determined by the scalar field $\tilde{\Sigma}_{s}$. Since we do
not use an iso-vector scalar field for the present parameter set NL3 the Dirac
mass is in these calculations identical for protons and neutrons. However,
this quantitiy should not be compared with the effective mass determined
empirically form a non-relativistic analysis of scattering data and of bound
states. From a non-relativistic approximation of the Dirac equations one finds
that the mass
\begin{equation}
m_{eff}=m-\tilde{\Sigma}_{0}%
\end{equation}
should be used for this purpose. Here $\tilde{\Sigma}_{0}$ is the time-like
component of the Lorentz vector field determined by the exchange of $\omega$-
and $\rho$-mesons.

In symmetric nuclear matter we find for NL3: $m_{D}/m=0.60$ and
$m_{eff}/m=0.67$. The latter value is smaller then the values
$m^{\ast}/m\approx0.71$ for protons and $m^{\ast}/m\approx0.76$ for
neutrons deduced from the calculated spectrum around the Fermi surface 
in simple RMF theory in Figs. \ref{f7} and \ref{f8}. 
Following similar arguments we would
obtain for RMF+PVC calculations an average effective mass of $0.89$.
This is obviously still too low as compared to the experimental
value.

From the other hand, around the Fermi surface where relativistic kinematic effects
are not significant our RMF+PVC spectrum can be characterized by the 
effective mass deduced from the Schr\"odinger equation which is a non-relativistic
limit of the Dirac equation (\ref{Deq1}). In this approximation one can calculate
the state-dependent E-mass ${\bar m}/m^{RMF}$ which is the inverted spectroscopic factor of the
dominant level $\lambda$:
\be
\frac{{\bar m}_k}{m^{RMF}} = \bigl [S^{(\lambda)}_k\bigr]^{-1}.
\ee 
For the calculated RMF+PVC spectrum the averaged E-masses are 1.26 for neutrons and
1.41 for protons if one takes into account all the states given in the Table \ref{tab3} 
with spectroscopic factors more then 0.5, i. e. good single-particle states.
Thus, the energy dependence of the mass operator increases the RMF neutron and proton effective
masses up to the values 0.96 and 1.0, respectively. 

\subsection{Comparison with other approaches}

\label{other} Although the problem of particle-vibration coupling in nuclei
has a long history and it was considered in a number of works, most of them
are based on a non-relativistic treatment of the nuclear many-body problem.
Only in a relatively recent investigation in Ref. \cite{VNR.02} a correction
of the RMF single-particle spectrum was undertaken in a phenomenological way
assuming a linear dependence of the mass operator near the Fermi surface. The
corresponding coupling constants were determined by a fit to nuclear ground
state properties. Despite the fact that the present approach is fully
microscopic without any additional parameter adjusted to experiment, we find
good agreement with the results of Ref. \cite{VNR.02} for the spectrum of
$^{208}$Pb. The shift caused by the phenomenological particle-vibrational
coupling in Ref. \cite{VNR.02} is only slightly larger than in the present investigation.

Non-relativistic microscopic investigations of particle-vibrational coupling
can be divided into two major groups. The first group
\cite{RW.73,KT.86,Pla.81,HS.76} uses a phenomenological single-particle input
to reproduce the experimental spectrum and has therefore to exclude the
contribution of the particle-vibration coupling from the full mass operator to
find the 'bare' spectrum. Usually these older approaches take into
consideration only a relatively small number of collective low-lying phonons
and use a particle-vibration coupling model \cite{BM.75}. This restriction to
only low-lying modes produces shifts less then 1 MeV. However, as it was shown
in Ref. \cite{HS.76}, enlarging of the phonon space with high-lying vibrations
leads to very strong shifts of the single-particle levels up to 4 MeV, and
no saturation is observed with respect to the dimension of the phonon space.

The second group of approaches (see, for instance Refs. \cite{PRS.80,BG.80})
starts from a \ self-consistent Hartree-Fock description and applies
perturbation theory to calculate the particle-vibration contribution to the
full mass operator. In such self-consistent methods it is more justified to
enlarge the phonon space. It was shown, for instance, in \cite{BG.80} that the
contribution of the isovector modes is noticeably smaller than the isoscalar
ones. The detailed investigation of the relative importance of the high
multipole states was performed in \cite{PRS.80}. Because of the larger phonon
space the typical shifts of the single-particle levels in $^{208}$Pb are about
1-2 MeV.

As for the spectroscopic factors, all the approaches predict similar values
because these factors are not very sensitive to the details of the
calculational schemes.

Thus one can see that the results of the present work are in a good agreement
with the results of earlier approaches.


\section{Summary}

The problem of the particle-vibration coupling is
considered on the foundation of the relativistic mean field approach. The Dyson
equation for the exact single-particle Green's function is solved in the Dirac
basis by taking into account the energy-dependent part of the fully
relativistic mass operator. This energy-dependent part is treated in terms of
the
particle-vibration coupling model that has been applied for the relativistic approach.

The particle-phonon coupling amplitudes have been computed within
self-consistent RRPA using the parameter set NL3 for the Lagrangian. A rather
large number of collective vibrational modes has been taken into account.
Relativistic effects of the Dirac sea on the mass operator have been analyzed
in the usual no-sea approximation. They have been found to be small as
compared to the contribution of the states with positive energy. Nontheless
the the Dirac sea contributions are crucial for the description of the RRPA
vibrations (Ref. \cite{RMG.01}) and are therefore fully taken into account.

Noticeable increase of the single-particle level density near the Fermi
surface relatively the pure RMF spectrum is obtained for $^{208}$Pb that
improves the agreement of the single-particle level scheme with experimental
data considerably. For four odd mass nuclei surrounding $^{208}$Pb the
distribution of the single-particle strength has been calculated and compared
with experiment as well as with the results obtained within several
non-relativistic approaches.

The major result of the present work is a consistent description of nuclear
many-body dynamics including complex configurations within an approach which
is (i) fully self-consistent, (ii) based on relativistic dynamics, (iii)
universally valid for nuclei all over the periodic table, and (iv) based on a
modern covariant density functional, which has been applied with great success
of many nuclear properties all over the periodic table. Complex configurations
play an important role in our understanding of the dynamics of the nuclear
many-body problem. Here we have discussed the single particle motion and its
coupling to collective vibrations. Such configurations are also of great
importance in the description of damping phenomena in even-even nuclei. Thus,
it is also interesting to investigate how the coupling to vibrational states
produces the spreading width within an extended relativistic RPA approach.
Work in this direction is in progress.

\leftline{\bf ACKNOWLEDGEMENTS}

We are thankful to Dr. V. I. Tselyaev and D. Pe\~{n}a Arteaga 
for useful comments and discussion. 
This work has been supported in part by the Bundesministerium f\"ur Bildung
und Forschung under project 06 MT 193. E. L. acknowledges the support from the
Alexander von Humboldt-Stiftung and the assistance and hospitality provided 
by the Physics Department of TU-M\"unchen. 


%
\end{document}